\documentclass[12pt,english,floatfix,nofootinbib,superscriptaddress,aps,prd,preprint]{revtex4}
\usepackage[utf8]{inputenc}
\usepackage{float}
\usepackage{array}
\usepackage{bbold}
\usepackage{lipsum}
\usepackage{dsfont}
\usepackage{graphicx}
\usepackage{amsmath,amsthm,amsfonts,amssymb}
\usepackage{graphicx}
\usepackage[english]{babel} 
\usepackage{color}
\usepackage{tensor}
\usepackage{esint}
\usepackage[dvips]{epsfig}
\usepackage[dvips]{graphicx}
\usepackage{float}
\usepackage{units}
\usepackage{textcomp}
\usepackage{mathrsfs}
\usepackage{amsmath}
\usepackage[makeroom]{cancel}
\usepackage{amssymb}
\usepackage{amsbsy}
\usepackage{amsfonts}
\usepackage{amssymb,mathrsfs,xcolor}
\usepackage{esint}
\usepackage{braket}
\usepackage{array}
\usepackage{graphicx}

\usepackage{wasysym}
\usepackage{multirow}
\usepackage{wrapfig}
\usepackage{subfig}

\usepackage{stmaryrd}
\usepackage{upgreek}

\makeatletter

\makeatletter\usepackage{babel}

\usepackage{hyperref}
\hypersetup{
    colorlinks,
    citecolor=blue,
    filecolor=green,
    linkcolor=purple,
    urlcolor=red,
}

\usepackage{slashed}

\newcommand{\ie}{\begin{equation}}
\newcommand{\fe}{\end{equation}}
\newcommand{\se}{\begin{eqnarray}}
\newcommand{\ff}{\end{eqnarray}}

\begin{document}

\title{An anisotropic rotating cosmic string with Lorentz violation: thermodynamics and Landau levels}

\author{A. A. Ara\'{u}jo Filho}
\email{dilto@fisica.ufc.br}

\affiliation{Departamento de Física, Universidade Federal da Paraíba, Caixa Postal 5008, 58051-970, João Pessoa, Paraíba,  Brazil.}

\author{J. A. A. S. Reis}
\email{jalfieres@gmail.com}

\affiliation{Universidade Estadual do Sudoeste da Bahia (UESB), Departamento de Ciências Exatas e Naturais, Campus Juvino Oliveira, Itapetinga -- BA, 45700-00,--Brazil}

\author{L. Lisboa--Santos}
\email{let\_lisboa@hotmail.com} 

\affiliation{Universidade Federal do Cear\'a (UFC), Departamento de F\'isica,\\ Campus do Pici,
Fortaleza - CE, C.P. 6030, 60455-760 - Brazil.}


\date{\today}

\begin{abstract}

In this work, we generalize the spacetime induced by a rotating cosmic string, taking into account anisotropic effects due the breaking of the Lorentz violation. In particular, we explore the energy levels of a massive spinless particle that is covariantly coupled to a uniform magnetic field aligned with the string. Subsequently, we introduce a scalar potential featuring both a Coulomb--type and a linear confining term and comprehensively solve the Klein–Gordon equations for each configuration. Finally, by imposing rigid--wall boundary conditions, we determine the Landau levels when the linear defect itself possesses magnetization. Notably, our analysis reveals the occurrence of Landau quantization even in the absence of gauge fields, provided the string possesses spin. Finally, the thermodynamic properties are computed as well in these scenarios.

\end{abstract}

\maketitle


\section{\protect\bigskip Introduction}

Over the past decade, cosmic strings have experienced a renaissance, marked by a resurgence of interest following a period of relative neglect \cite{01,02,03,04,05,06,07}. Cosmic strings, theoretical massive entities, are conjectured to have played a potential albeit marginal role in influencing the anisotropy observed in the cosmic microwave background radiation. This influence, in turn, could contribute to shaping the large-scale structure of the universe \cite{08}. Intriguingly, their existence finds theoretical support in superstring theories that encompass scenarios with either compactified or extended extra dimensions.

Both static and rotating cosmic strings emerge as equally compelling contributors to several notable effects. These effects encompass the exertion of particle self-force \cite{09,10}, gravitational lensing \cite{11}, and the generation of highly energetic particles \cite{12,13,14}. This renewed exploration into the properties and consequences of cosmic strings not only revitalizes their role in cosmological discussions but also underscores their potential impact on our understanding of the universe.

Rotating cosmic strings, along with their static counterparts, manifest as one-dimensional stable topological defects, likely formed during the early stages of the universe. Their distinct characteristics are encapsulated by a wedge parameter, denoted as $\alpha$, contingent on the linear mass density $\mu$, and the linear density of angular momentum $J$. Originally conceptualized as general relativistic solutions within a Kerr spacetime in ($1 + 2$) dimensions \cite{15}, these descriptions were subsequently naturally extended to encompass the four--dimensional spacetime \cite{16}.

Remarkably, beyond the singularity, cosmic strings--whether static or rotational--exhibit a flat spacetime geometry, endowed with noteworthy global properties. Among these properties are theoretically predicted effects such as gravitomagnetism and the (non--quantum) gravitational Aharonov--Bohm effect \cite{17,18}. This dual nature of cosmic strings, as stable topological defects and as contributors to gravitational phenomena, underscores their profound implications for our understanding of the cosmos.

Cosmic strings, as postulated in theoretical physics, might exhibit an internal structure, as suggested in prior research \cite{20}. This internal structure can give rise to a Gödel spacetime configuration characterized by an exotic region that permits the existence of closed timelike curves (CTCs) around the singularity. Notably, the boundary of this region is situated at a distance proportional to $J/\alpha$ from the cosmic string, thereby providing a natural boundary condition.

Rotating cosmic strings have also been extensively investigated within the framework of the Einstein--Cartan theory \cite{21,22} and teleparallel gravity \cite{23}. In these studies, particular attention has been devoted to examining the region where CTCs manifest. Additionally, investigations into cosmic strings have extended into the realm of extra dimensional theories, encompassing considerations of their causal structure. However, it is essential to note that these extra dimensional studies have prompted criticisms regarding the realistic existence of the region accommodating CTCs \cite{24}.

Cosmic strings, as posited in theoretical physics, may harbor an internal structure, as suggested in recent investigations \cite{20}. This internal configuration has the intriguing consequence of giving rise to a Gödel spacetime, characterized by an exotic region that facilitates the existence of CTCs around the singularity. Significantly, the boundary of this CTC--permitting region is situated at a distance proportional to $J/\alpha$ from the cosmic string, providing a natural and consequential boundary condition.

The phenomenon of rotating cosmic strings has been subject to in--depth exploration within the frameworks of both the Einstein--Cartan theory \cite{21,22} and teleparallel gravity \cite{23}.  Furthermore, the investigation of cosmic strings extends into the realm of extra-dimensional theories, exploring aspects of their causal structure. However, it is noteworthy that these extra-dimensional studies have not been without critique, particularly regarding doubts about the authentic existence of the CTCs \cite{24}.

The exploration of Landau levels within the spacetime of a stationary spinning cosmic string has been relatively scarce in the literature, as evidenced by limited references \cite{25,26}. In stark contrast, the literature on static cosmic strings is more extensive and well-documented, as seen in numerous works \cite{27,28,29,30} and their associated references. This apparent disparity may stem from the considerable focus on static cosmic strings, driven by analogies and potential technological applications in condensed matter physics, such as disclination phenomena in crystals \cite{31}.

Investigations into the thermodynamics of Lorentz--violating extensions across diverse field theory models offer valuable insights into the early stages of the Universe's expansion. This period, characterized by sizes commensurate with the scales of Lorentz symmetry breaking \cite{kostelecky2011data}, is crucial for understanding fundamental aspects of the cosmos. The methodology for examining thermodynamic aspects within LV theories was initially introduced in \cite{colladay2004statistical}, marking a pivotal starting point. Since then, the application of this approach has seen development across various studies \cite{casana2008lorentz,casana2009finite,gomes2010free,aa2020lorentz,araujo2022particles,araujo2021bouncing,maluf2020thermodynamic,reis2020thermal,das2009relativistic,araujo2021higher,araujo2022thermal,araujo2023thermodynamics}.

In this study, we extend the analysis of the spacetime induced by a rotating cosmic string, considering anisotropic effects stemming from the breaking of Lorentz invariance. Specifically, we investigate the energy levels of a massive spinless particle that is covariantly coupled to a uniform magnetic field aligned with the string. Subsequently, we introduce a scalar potential comprising both a Coulomb--type and a linear confining term, thoroughly solving the Klein--Gordon equations for each configuration. By imposing rigid--wall boundary conditions, we ascertain the Landau levels in instances where the linear defect itself possesses magnetization. Significantly, our analysis uncovers the occurrence of Landau quantization even in the absence of gauge fields, provided the cosmic string possesses spin. Lastly, we compute the thermodynamic properties in these scenarios, providing a comprehensive exploration of the interplay between the cosmic string's rotation, anisotropic effects, and the quantum behavior of particles within this spacetime.


\section{Spinless charged particle}

Initiating our examination, let us explore the characteristics of a massive, charged, relativistic spinless quantum particle within the spacetime framework of an idealized stationary rotating cosmic string. Notably, the term ``stationary'' denotes the absence of structural intricacies in the cosmic string, and its metric is precisely articulated as follows \cite{mazur1986spinning}:
\begin{equation}
ds^{2}=c^{2}dt^{2}+2acdtd\phi -\left( \alpha ^{2}\rho ^{2}-a^{2}\right)
d\phi ^{2}-d\rho ^{2}-dz^{2}.  \label{1}
\end{equation}%
In this configuration, the cosmic string is positioned along the $z$--axis, and the cylindrical coordinates are denoted as $\left( t, \rho, \phi, z \right)$ with standard ranges. The rotation parameter, $a$, is defined as $4GJ/c^{3}$, where $G$ stands for the gravitational Newton constant and has units of distance. The wedge parameter, $\alpha$, determining the angular deficit $\Delta \phi = 2\pi \left( 1-\alpha \right)$ induced by the cosmic string, is given by $\alpha = 1-4\mu G/c^{2}$. Here, $c$ represents the speed of light, $G$ is the gravitational Newton constant, and $\mu$ is the linear density of the mass of the string.

To explore relativistic quantum motion within a curved spacetime and in the presence of a gauge potential, we turn our attention to the covariant form of the Klein-Gordon equation, expressed as:
\begin{equation}
\left[ \frac{1}{\sqrt{-g}}D_{\mu }\left( \sqrt{-g}g^{\mu \nu }D_{\nu
}\right) +\frac{m^{2}c^{2}}{\hbar ^{2}}\right] \Psi =0,  \label{2}
\end{equation}
where, $D_{\mu }=\partial _{\mu }-\frac{ie}{\hbar c}A_{\mu }$, where $e$ signifies the electric charge, and $m$ denotes the particle's mass. The Planck constant is represented by $\hbar$, $g^{\mu \nu }$ is the metric tensor, and $g=\det g^{\mu \nu }$. Assuming the presence of a magnetic field parallel to the cosmic string, the vector potential can be defined as:
\begin{equation*}
\mathbf{A}=\left( 0,A_{\phi },0\right) ,
\end{equation*}%
where $A_{\phi }=\frac{1}{2}\alpha \mathcal{K}_{\phi j}B^{j}\rho =\frac{1}{2}%
\alpha \mathcal{K}_{\phi }\cdot \mathbf{B}\rho $, and $K_{\phi j}$ represents the components of a constant tensor that introduces 

The cylindrical symmetry inherent in the background space, as described by Eq. (\ref{1}), implies the possibility of factorizing the solution to Eq. (\ref{2}) as:
\begin{equation}
\Psi =\left( \rho ,\phi ,z,t\right) =e^{-i\frac{E}{\hbar }t}e^{i\left( l\phi
+k_{z}z\right) }R\left( \rho \right) .  \label{3}
\end{equation}%
Here, $R\left( \rho \right)$ denotes the solution to the radial equation, characterized by:
\begin{equation}
\frac{d^{2}R}{d\rho ^{2}}+\frac{1}{\rho }\frac{dR}{d\rho }-\Lambda \frac{R}{%
\rho ^{2}}-\frac{e^{2}\left( \mathcal{K}_{\phi }\cdot \mathbf{B}\right)
^{2}\rho ^{2}}{4\hbar ^{2}c^{2}}R+\Delta R=0,  \label{4}
\end{equation}%
in which%
\begin{eqnarray}
\Lambda  &=&\left( \frac{l}{\alpha }+\frac{aE}{\alpha \hbar c}\right) ^{2},
\label{5} \\
\Delta  &=&\frac{E^{2}}{\hbar ^{2}c^{2}}-\frac{m^{2}c^{2}}{\hbar ^{2}}%
-k_{z}^{2}+\frac{e\mathcal{K}_{\phi }\cdot \mathbf{B}}{\hbar c}\left( \frac{l%
}{\alpha }+\frac{aE}{\alpha \hbar c}\right) ,  \label{6}
\end{eqnarray}%
with $k_{z}$ and $E$ represent the $z$-momentum and energy of the particle, respectively, while $l$ denotes the azimuthal angular quantum number. The solutions to Eq. (\ref{4}) can be determined through the application of the following transformation:%
\begin{equation}
R\left( \rho \right) =\exp \left( -\frac{e\rho ^{2}\mathcal{K}_{\phi }\cdot 
\mathbf{B}}{4\hbar c}\right) \rho ^{\sqrt{\Lambda }}F\left( \rho \right) .
\end{equation}%

Upon substituting the aforementioned expression into Eq. (\ref{4}), we acquire:
\begin{equation}
\rho F^{^{\prime \prime }}\left( \rho \right) +\left( 1+2\sqrt{\Lambda }-%
\frac{e\rho ^{2}\mathcal{K}_{\phi }\cdot \mathbf{B}}{\hbar c}\right)
F^{^{\prime }}\left( \rho \right) +\left[ \Delta -\frac{e\mathcal{K}_{\phi
}\cdot \mathbf{B}}{\hbar c}\left( 1+\sqrt{\Lambda }\right) \right] \rho
F\left( \rho \right) =0.  \label{8}
\end{equation}%
Presently, let us contemplate the transformation of variables
\begin{equation*}
z=\frac{e\mathcal{K}_{\phi }\cdot \mathbf{B}}{2\hbar c}\rho ^{2}.
\end{equation*}%
Therefore, Eq. (\ref{8}) takes on the customary form:
\begin{equation}
zF^{^{\prime \prime }}\left( z\right) +\left( 1+\sqrt{\Lambda }-z\right)
F^{^{\prime }}\left( z\right) -\left[ \frac{1}{2}\left( 1+\sqrt{\Lambda }%
\right) -\frac{\hbar c}{2e\mathcal{K}_{\phi }\cdot \mathbf{B}}\Delta \right]
F\left( z\right) =0.
\end{equation}%

This is the renowned confluent hypergeometric equation, and its linearly independent solutions are:
\begin{equation}
F^{\left( 1\right) }\left( z\right) =_{1}F_{1}\left( \frac{1}{2}+\frac{\sqrt{%
\Lambda }}{2}-\frac{\hbar c}{2e\mathcal{K}_{\phi }\cdot \mathbf{B}}\Delta ;%
\text{ }1+\sqrt{\Lambda };\text{ }z\right) ,
\end{equation}%
\begin{equation}
F^{\left( 2\right) }\left( z\right) =z^{-\sqrt{\Lambda }}{}_{1}F_{1}\left( 
\frac{1}{2}-\frac{\sqrt{\Lambda }}{2}-\frac{\hbar c}{2e\mathcal{K}_{\phi
}\cdot \mathbf{B}}\Delta ;\text{ }1-\sqrt{\Lambda };\text{ }z\right) .
\end{equation}%

Thus, the radial solutions, denoted as $R\left( \rho \right)$, can be expressed as:
\begin{equation}
R^{\left( 1\right) }\left( \rho \right) =A_{1}\exp \left( -\frac{e\rho ^{2}%
\mathcal{K}_{\phi }\cdot \mathbf{B}}{4\hbar c}\right) \rho ^{\sqrt{\Lambda }%
}{}_{1}F_{1}\left( \frac{1}{2}+\frac{\sqrt{\Lambda }}{2}-\frac{\hbar c}{2e%
\mathcal{K}_{\phi }\cdot \mathbf{B}}\Delta ;\text{ }1+\sqrt{\Lambda };\text{ 
}\frac{e\rho ^{2}\mathcal{K}_{\phi }\cdot \mathbf{B}}{2\hbar c}\right) ,
\end{equation}%
\begin{equation}
R^{\left( 2\right) }\left( \rho \right) =A_{2}\exp \left( -\frac{e\rho ^{2}%
\mathcal{K}_{\phi }\cdot \mathbf{B}}{4\hbar c}\right) \rho ^{-\sqrt{\Lambda }%
}{}_{1}F_{1}\left( \frac{1}{2}-\frac{\sqrt{\Lambda }}{2}-\frac{\hbar c}{2e%
\mathcal{K}_{\phi }\cdot \mathbf{B}}\Delta ;\text{ }1-\sqrt{\Lambda };\text{ 
}\frac{e\rho ^{2}\mathcal{K}_{\phi }\cdot \mathbf{B}}{2\hbar c}\right).
\end{equation}%
In this context, let $A_{1}$ and $A_{2}$ denote normalization constants. The second solution, found to be unsuitable at the origin, is consequently excluded. Due to the exponential divergence of confluent hypergeometric functions as $\rho$ tends to infinity, the condition that must be enforced for physically acceptable solutions is:%
\begin{equation}
\frac{1+\sqrt{\Lambda }}{2}-\frac{\hbar c}{2e\mathcal{K}_{\phi }\cdot 
\mathbf{B}}\Delta =-n. \label{14}
\end{equation}%
Here, $n$ is a positive integer. By substituting the expressions for $\Lambda$ and $\Delta$ from Eqs. (\ref{5}) and (\ref{6}) into Eq. (\ref{14}), the resulting expression is as follows:%
\begin{equation}
\frac{E^{2}}{e\hbar c\mathcal{K}_{\phi }\cdot \mathbf{B}}+\left( \frac{l}{%
\alpha }+\frac{aE}{\alpha \hbar c}\right) -\left\vert \frac{l}{\alpha }+%
\frac{aE}{\alpha \hbar c}\right\vert -\frac{c}{e\hbar \mathcal{K}_{\phi
}\cdot \mathbf{B}}\left( \hbar ^{2}k^{2}+m^{2}c^{2}\right) -\frac{1}{4}\frac{%
ea^{2}\mathcal{K}_{\phi }\cdot \mathbf{B}}{\hbar c^{3}}=2n+1.  \label{15}
\end{equation}%

After that, we are able to address the energy eigenvalues as%
\begin{eqnarray}
E_{n,l} &=&\frac{ea\mathcal{K}_{\phi }\cdot \mathbf{B}}{2\alpha }\left( 
\frac{\left\vert l\right\vert -l}{l}\right)   \notag \\
&&\pm \sqrt{m^{2}c^{4}+\hbar ^{2}k^{2}c^{2}+\left( \frac{ea}{2\alpha }\frac{%
\left\vert l\right\vert -l}{l}\mathcal{K}_{\phi }\cdot \mathbf{B}\right)
^{2}+\hbar ce\left( 2n+1+\frac{\left\vert l\right\vert }{\alpha }-\frac{l}{%
\alpha }\right) \mathcal{K}_{\phi }\cdot \mathbf{B}}.  \label{16}
\end{eqnarray}%

This expression highlights the lack of invariance in the energy eigenvalues under the interchange of negative and positive eigenvalues of the azimuthal quantum number $l$. Such non--invariance is a consequence of the spacetime's topological twist around the rotating string, dependent not only on $\alpha$ but also on $a$ (refer to Eq. (\ref{1})). It is noteworthy that by deactivating the string's rotation, i.e., setting $a=0$, the resulting expression aligns with a previously established formulation \cite{cunha2016relativistic,figueiredo2012relativistic}, applicable to the static string. Additionally, it is observed that for positive values of $l$, the energy spectra for both static and rotating strings remain identical.


\subsection{Non--relativistic limit}

The non--relativistic expression can be derived through consideration of:%
\begin{equation*}
\frac{E^{2}}{c^{2}}-m^{2}c^{2}\approx 2mE.
\end{equation*}%
In the preceding equation, under these conditions, Eq. (\ref{16}) transforms into:
\begin{equation}
E_{n,l}\approx \frac{1}{1+\frac{ea\mathcal{K}_{\phi }\cdot \mathbf{B}}{%
2\alpha mc^{2}}\left( 1-\frac{\left\vert l\right\vert }{l}\right) }\left[ 
\frac{\hbar ^{2}k^{2}}{2m}+\frac{e\hbar }{2mc}\left( 2n+1+\frac{\left\vert
l\right\vert }{\alpha }-\frac{l}{\alpha }\right) \mathcal{K}_{\phi }\cdot 
\mathbf{B}\right] .
\end{equation}%
Consequently, it becomes apparent that when $l > 0$ (indicating a particle orbiting parallel to the string's rotation), the energy levels remain consistent for both static \cite{de2001landau} and rotating strings. However, for anti parallel orbits ($l < 0$), the permitted spectrum is contingent upon the angular momentum density of the string.

In this scenario, when employing the slow rotation approximation, wherein terms of $\mathcal{O}\left( a^{2}\right)$ are disregarded, we obtain:
\begin{equation}
\Delta E_{n,l}/E_{n,l}^{\left( 0\right) }\approx -\frac{ea}{\alpha mc^{2}}%
\mathcal{K}_{\phi }\cdot \mathbf{B}.
\end{equation}%
Here, $\Delta E_{n,l}$ represents the relative difference between our obtained result and $E_{n,l}^{\left( 0\right) }$, corresponding to the levels of the static string. This outcome represents an enhancement compared to the results presented in \cite{cunha2016relativistic}, where additional approximations were introduced.


\subsection{Thermodynamics properties}

Utilizing the canonical ensemble enables the examination of various thermodynamic properties. Specifically, the free energy exhibits a monotonically decreasing trend with temperature, while the entropy displays a monotonically increasing pattern. It is noteworthy that an escalation in the magnetic field leads to a reduction in the values of the aforementioned thermodynamic functions, including internal energy. Conversely, the heat capacity experiences an increase within the range of $0\leq T\leq 5~\mathrm{eV}$. Beyond $T>5~\mathrm{eV}$, an inversion in this behavior becomes apparent.
\begin{figure}[tbh]
\centering
  \subfloat[Free Energy]{\includegraphics[width=8cm,height=5cm]{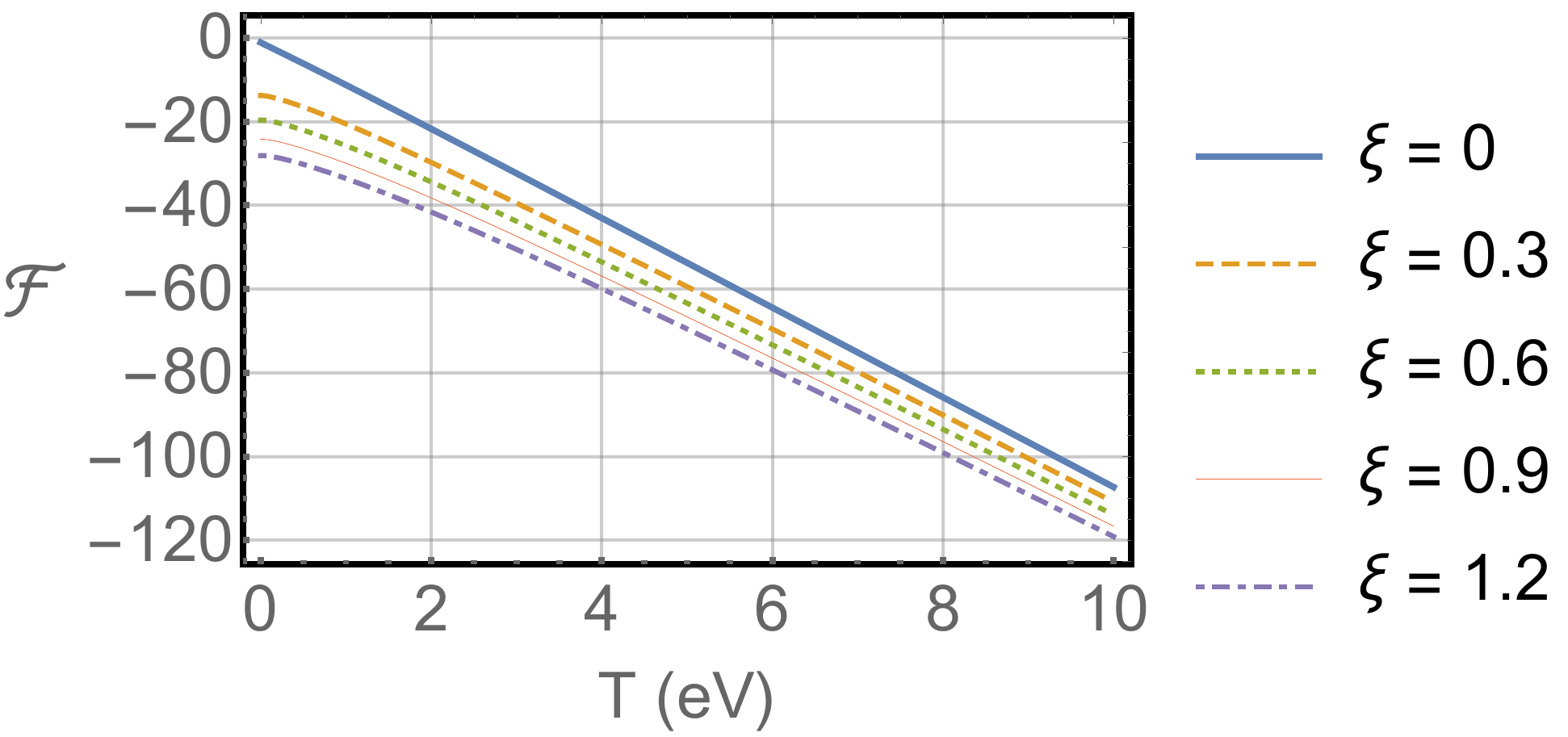}
  \label{fig:GPCase-2-lowTN}}
  \subfloat[Internal energy]{\includegraphics[width=8cm,height=5cm]{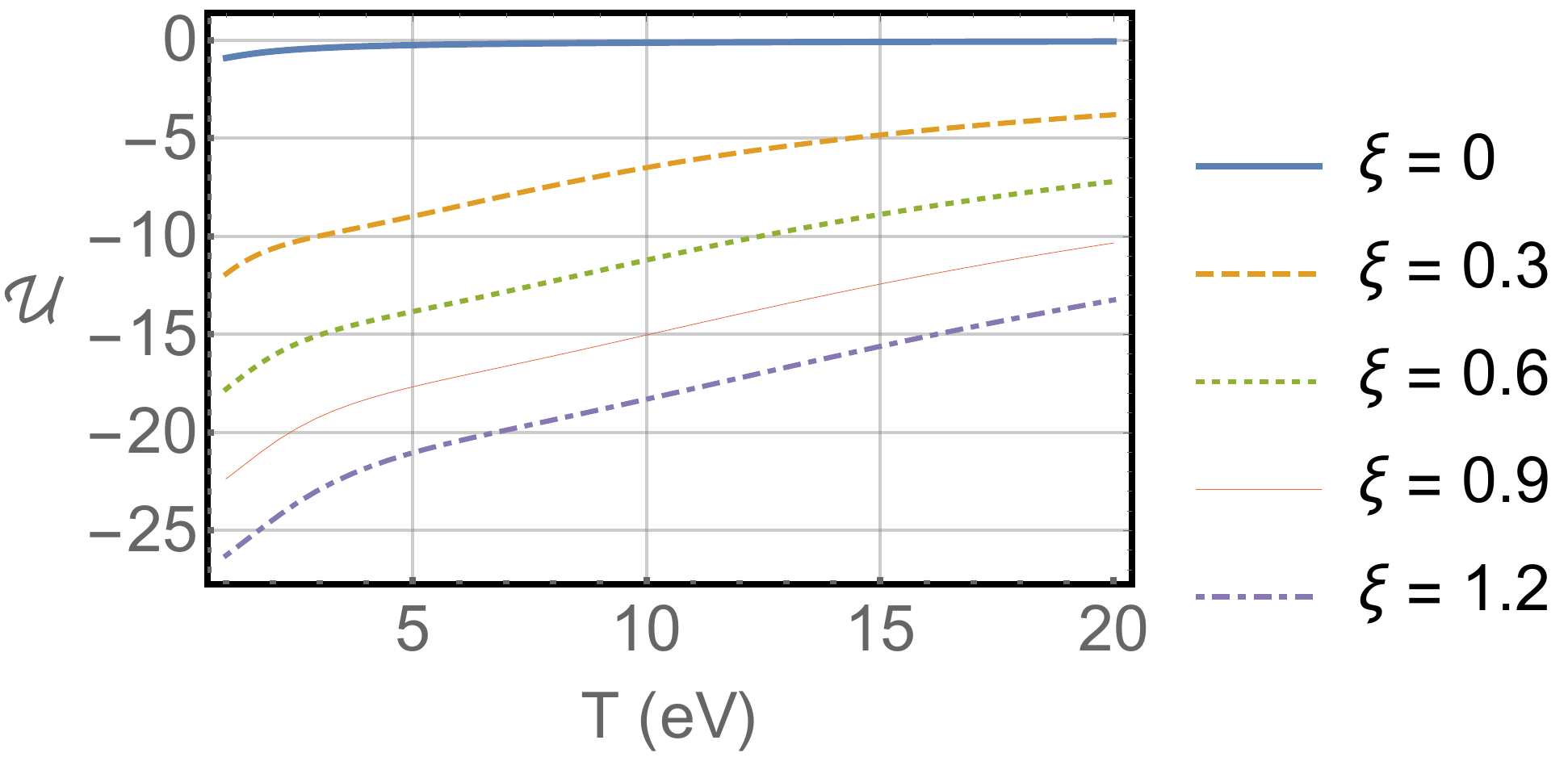}
  \label{fig:GPCase-2-lowTU}}\\
  \subfloat[Entropy]{\includegraphics[width=8cm,height=5cm]{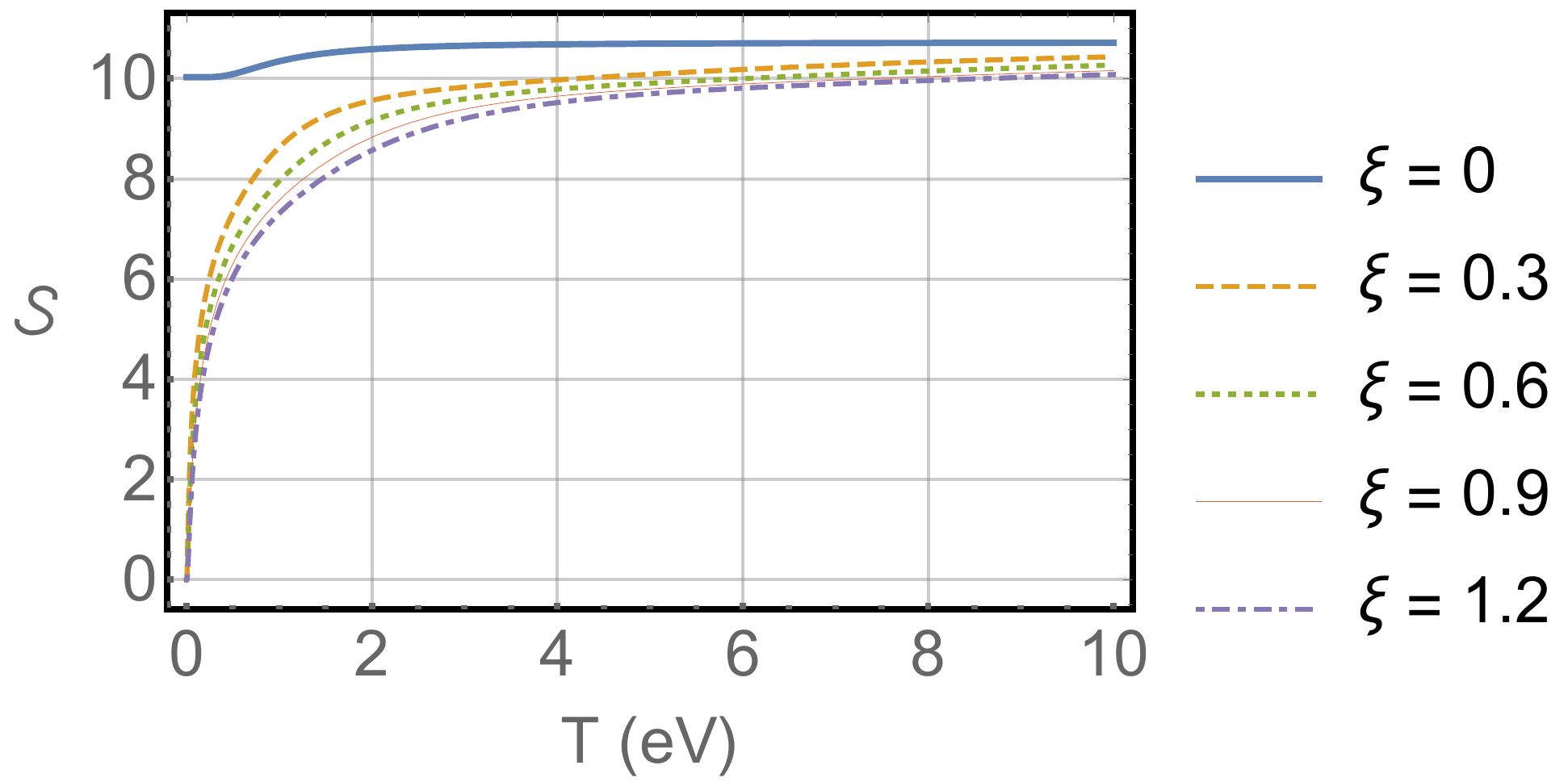}
  \label{fig:GPCase-2-lowTS}}
  \subfloat[Heat capacity]{\includegraphics[width=8cm,height=5cm]{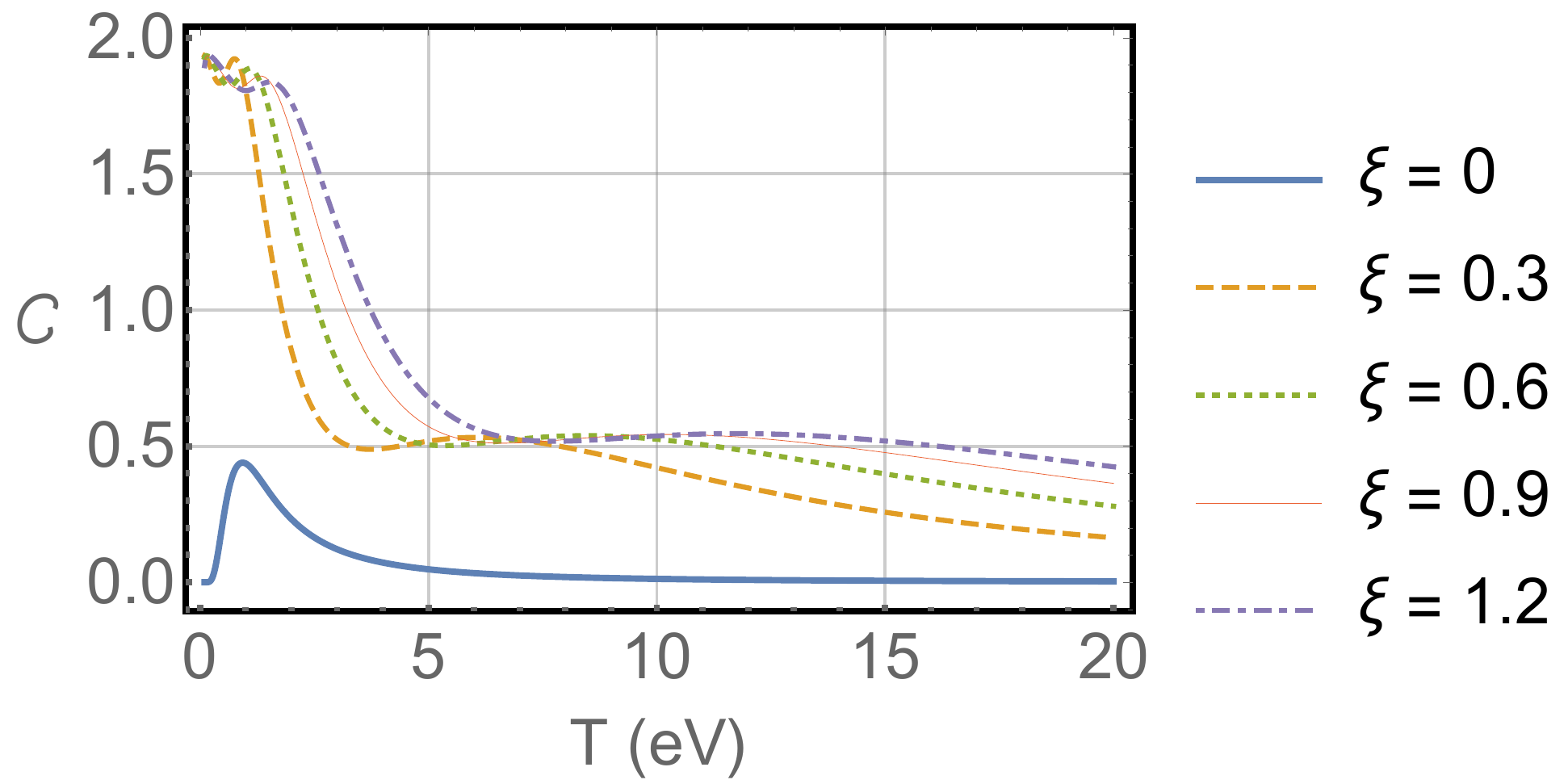}
  \label{fig:GPCase-2-lowTC}}
 \caption{Free Energy, internal energy, entropy and heat capacity are displayed for different values of $\xi$.}
\end{figure}

At a temperature of $T=1~\mathrm{eV}$, we discern a phase transition phenomenon characterized by the presence of a mound. This behavior is mitigated with the elevation of the external magnetic field.

\begin{figure}[tbh]
\centering
  \subfloat[Permittivity]{\includegraphics[width=8cm,height=5cm]{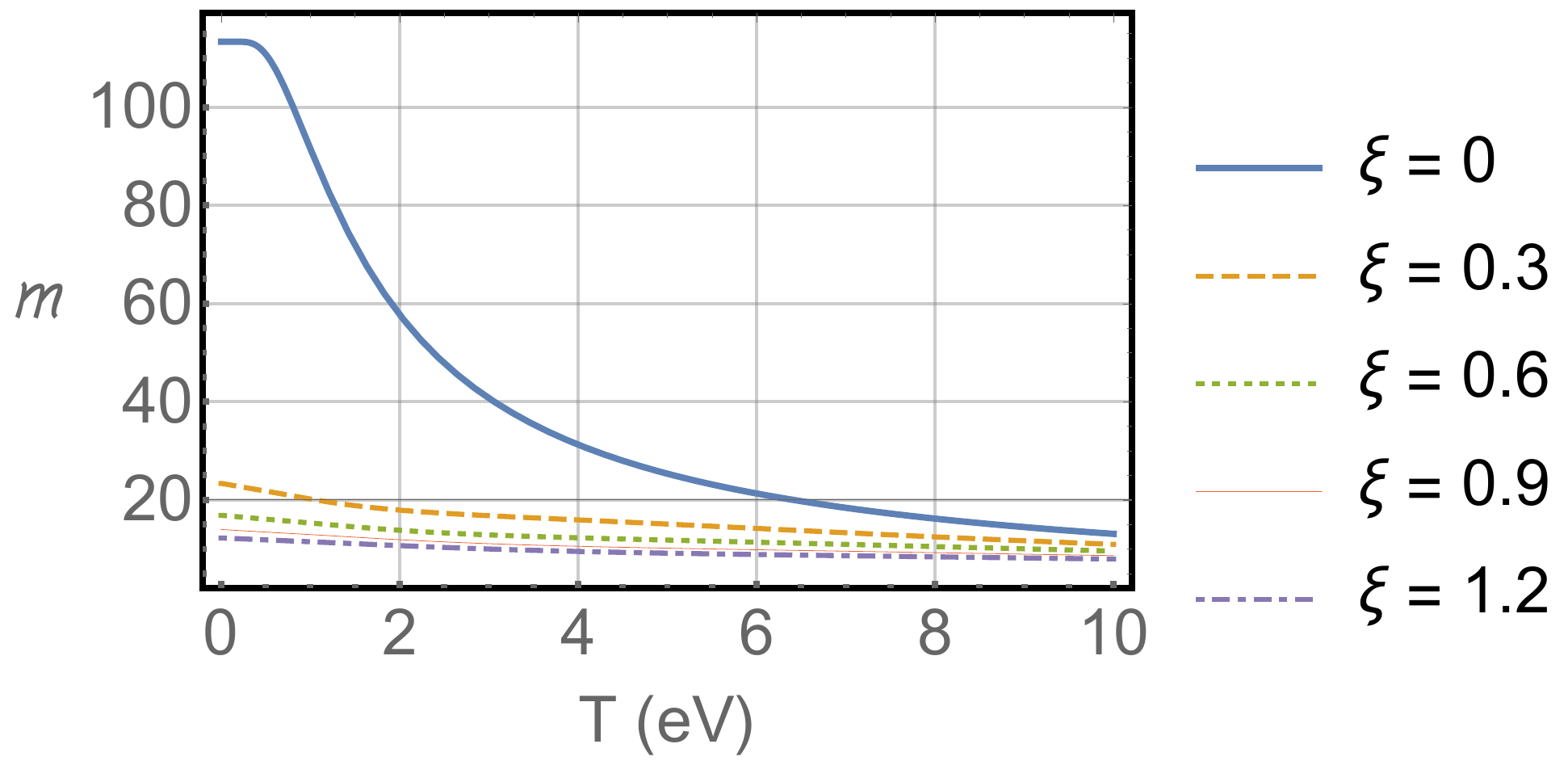}
  \label{fig:GPCase-2-lowTm}}
  \subfloat[Susceptibility]{\includegraphics[width=8cm,height=5cm]{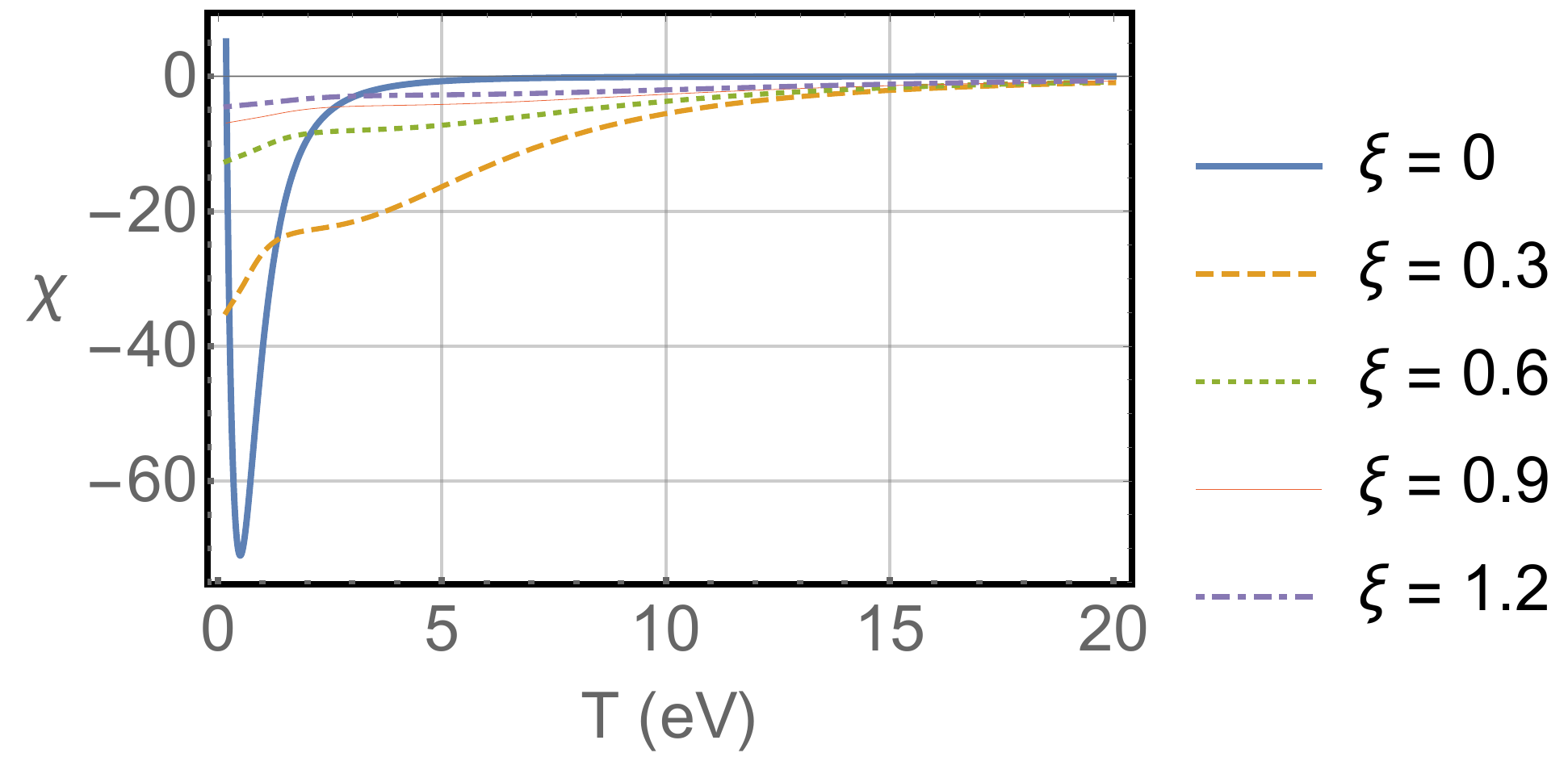}
  \label{fig:GPCase-2-lowTx}}
  \caption{Permittivity and susceptibility are shown for different values of $\xi$}
\end{figure}

Additionally, we compute the permittivity and susceptibility of the system. It is noteworthy that, in both instances, we reach a saturated value for high temperatures, indicating that the magnetic field plays no significant role in these conditions. Nonetheless, for temperatures below $15~\mathrm{eV}$, we note a magnetic dependence within the system. Interestingly, the susceptibility exhibits a decrease with the escalation of the magnetic field, accompanied by a similarly intriguing behavior of decreasing magnetization. In addition, it is important to mention that thermodynamic functions were also calculated to different geometries \cite{aa,bb,cc,dd}.


\section{Cylindrically Symmetric Scalar Potential}

In this section, we aim to extend the analysis presented earlier by introducing a novel element—a distinctive anisotropic scalar potential
\begin{equation}
S_{m}\left( \rho \right) =\frac{\boldsymbol{\kappa }\cdot \boldsymbol{r}}{%
\rho ^{2}}+\boldsymbol{\nu }\cdot \boldsymbol{r},  \label{19}
\end{equation}%
where $\boldsymbol{\kappa}$ and $\boldsymbol{\nu}$ are constant vectors introducing anisotropies, $\boldsymbol{r}=\rho \boldsymbol{\hat{\rho}}%
+\rho \boldsymbol{\hat{\varphi}}+\boldsymbol{\hat{z}}$ is a vector in
cylindrical coordinates. To assess the impact of this potential on the quantum dynamics of the particle, $\boldsymbol{\kappa }%
=\left( 0,\kappa _{\varphi },0\right) $ and $\boldsymbol{\nu }=\left( 0,\nu
_{\varphi },0\right) $, we need to change Eq.$\left( \ref{2}\right) $ by
considering Eq.$\left( \ref{19}\right) $ to the mass term so that $%
\frac{mc}{\hbar }$\ is changed by $\frac{mc}{\hbar }+S_{m}\left( \rho
\right)$. Introducing this modification into Eq. $\left( \ref{2} \right)$ and adopting the ansatz from Eq. $\left( \ref{3} \right)$, we derive the subsequent radial equation:%
\begin{equation}
\frac{d^{2}R}{d\rho ^{2}}+\frac{1}{\rho }\frac{dR}{d\rho }-\mathcal{F}\frac{R%
}{\rho ^{2}}-2M\kappa _{\varphi }\frac{R}{\rho }-2MR\nu _{\varphi }\rho
-\Omega ^{2}\rho ^{2}R+\mathcal{D}R=0,  \label{20}
\end{equation}%
in which%
\begin{eqnarray}
M &=&\frac{mc}{\hbar }, \\
\Omega ^{2} &=&M^{2}\omega ^{2}-\nu _{\varphi }^{2}, \\
\mathcal{F} &\mathcal{=}&\left( \frac{l}{\alpha }+\frac{a}{\alpha }%
\varepsilon \right) ^{2}+\kappa _{\varphi }^{2}, \\
\mathcal{D} &\mathcal{=}&\varepsilon ^{2}+2M\omega \left( \frac{l}{\alpha }+%
\frac{a}{\alpha }\varepsilon \right) -M^{2}-2\kappa _{\varphi }\nu _{\varphi
}-k_{z}^{2},
\end{eqnarray}%
and%
\begin{eqnarray*}
2M\omega  &=&\frac{e}{\hbar c\alpha }\mathcal{K}_{\phi }\cdot \mathbf{B}, \\
\varepsilon  &=&\frac{E}{\hbar c}.
\end{eqnarray*}%
For ease of notation, let us introduce a new function $H\left(\rho\right)$ as follows:
\begin{equation}
R\left( \rho \right) =\exp \left( -\frac{1}{2}\Omega \rho ^{2}-\frac{M\nu
_{\varphi }}{\Omega }\rho \right) \rho ^{\sqrt{\mathcal{F}}}H\left( \rho
\right) .  \label{25}
\end{equation}%
In this sense, we can use the redefinition $\sqrt{\Omega }\rho \rightarrow \rho $, Eq.$%
\left( \ref{20}\right) $, leading to%
\begin{gather}
\frac{d^{2}H}{d\rho ^{2}}+\left( \frac{1+2\sqrt{\mathcal{F}}}{\rho }-\frac{%
2M\nu _{\varphi }}{\Omega ^{3/2}}-2\rho \right) \frac{dH}{d\rho }  \notag \\
+\left[ \frac{M^{2}\nu _{\varphi }^{2}}{\Omega ^{3}}+\frac{\mathcal{D}}{%
\Omega }-2\sqrt{\mathcal{F}}-2-\frac{1}{2}\left( \frac{4M\kappa _{\varphi }}{%
\sqrt{\Omega }}+\left( 1+2\sqrt{\mathcal{F}}\right) \frac{2M\nu _{\varphi }}{%
\Omega ^{3/2}}\right) \frac{1}{\rho }\right] H=0,  \label{26}
\end{gather}%
or
\begin{equation}
H_{b}^{^{{}}\prime \prime }\left( z\right) +\left( \frac{1+\alpha }{z}-\beta
-2z\right) H_{b}^{^{{}}\prime }\left( z\right) +\left[ \gamma -\alpha -2-%
\frac{1}{2}\left[ \delta +\left( 1+\alpha \right) \beta \right] \frac{1}{z}%
\right] H_{b}\left( z\right) =0.  \label{27}
\end{equation}%
The solution of above equation is the biconfluent Heun functions given below
\begin{equation}
H_{b}\left( z\right) =C_{1}H_{b}\left( \alpha ,\beta ,\gamma ,\delta
;z\right) +C_{2}z^{-\alpha }H_{b}\left( -\alpha ,\beta ,\gamma ,\delta
;z\right) ,
\end{equation}%
where $C_{1}$\ and $C_{2}$\ are simply the normalization constants. When $\alpha $\ is
a positive integer,  we have \cite{vieira2015quantum}%
\begin{equation}
H_{b}\left( \alpha ,\beta ,\gamma ,\delta ;z\right) =\sum_{j=0}^{\infty }%
\frac{A_{j}}{\left( 1+\alpha \right) _{j}}\frac{z^{j}}{j!},
\end{equation}%
with coefficients $A_{j}$ follow the three--term recurrence relation: $%
\left( j\geq 0\right) $,%
\begin{equation}
A_{j+2}=\left[ \left( j+1\right) \beta +\frac{1}{2}\left[ \delta +\left(
1+\alpha \right) \beta \right] \right] A_{j+1}-\left( j+1\right) \left(
j+1+\alpha \right) \left( \gamma -\alpha -2-2j\right) A_{j}.
\end{equation}%

After taking a compassion of Eqs.$\left( \ref{26}\right) $ and $\left( \ref{27}\right) 
$, we get:%
\begin{equation}
H^{\left( 1\right) }\left( \rho \right) =c_{1}H_{b}\left( 2\sqrt{\mathcal{F}}%
,\frac{2M\nu }{\Omega ^{3/2}},\frac{M^{2}\nu ^{2}}{\Omega ^{3}},\frac{%
\mathcal{D}}{\Omega },\frac{4M\kappa }{\sqrt{\Omega }};\sqrt{\Omega }\rho
\right) ,  \label{31}
\end{equation}%
\begin{equation}
H^{\left( 2\right) }\left( z\right) =c_{2}\rho ^{-2\sqrt{\mathcal{F}}%
}H_{b}\left( -2\sqrt{\mathcal{F}},\frac{2M\nu }{\Omega ^{3/2}},\frac{%
M^{2}\nu ^{2}}{\Omega ^{3}},\frac{\mathcal{D}}{\Omega },\frac{4M\kappa }{%
\sqrt{\Omega }};\sqrt{\Omega }\rho \right).  \label{32}
\end{equation}%
Notice that upon substituting $\rho \rightarrow \sqrt{\Omega }\rho$ into the preceding expressions, we note that the solution derived from Eq. $\left( \ref{32} \right)$ diverges at the origin. Consequently, we dismiss this solution. Furthermore, given the high divergence of the biconfluent Heun functions at infinity, our attention must shift towards their polynomial forms. Notably, the biconfluent Heun function assumes a polynomial of degree $n$ if the following conditions are simultaneously met:%
\begin{eqnarray}
\gamma -\alpha -2 &=&2n,\text{ \ \ \ \ \ \ \ \ }n=0,1,2,3,...,  \label{33} \\
A_{n+1} &=&0.  \label{34}
\end{eqnarray}%
Here, $A_{n+1}$ has $n+1$ real roots when $1+\alpha >0$ and $\beta \in
\mathbb{R}
$. It is denoted as a tridiagonal $\left( n+1 \right)$--dimensional determinant
\begin{equation}
\left\vert 
\begin{array}{ccccccc}
\delta ^{\prime } & 1 & 0 & 0 & \cdots  & \cdots  & 0 \\ 
2\left( 1+\alpha \right) n & \delta ^{\prime }-\beta  & 1 & 0 & \cdots  & 
\cdots  & 0 \\ 
0 & 4\left( 2+\alpha \right) \left( n-1\right)  & \delta ^{\prime }-2\beta 
& 1 & 0 & \cdots  & 0 \\ 
0 & 0 & \gamma _{2} & \delta ^{\prime }-3\beta  & 1 & \cdots  & \vdots  \\ 
\vdots  & \vdots  & 0 & \ddots  & \ddots  & \ddots  & 0 \\ 
\vdots  & \vdots  & \vdots  & \vdots  & \gamma _{j-1} & \delta
_{s-1}^{\prime } & 1 \\ 
0 & 0 & 0 & 0 & 0 & \gamma _{s} & \delta _{s}^{\prime }%
\end{array}%
\right\vert =0,  \label{35}
\end{equation}%
with%
\begin{eqnarray}
\delta ^{\prime } &=&-\frac{1}{2}\left[ \delta +\left( 1+\alpha \right)
\beta \right] ,  \label{36} \\
\gamma _{s} &=&2\left( s+1\right) \left( s+1+\alpha \right) \left(
n-s\right) ,  \label{37} \\
\delta _{s}^{\prime } &=&\delta ^{\prime }-\left( s+1\right) \beta .
\label{38}
\end{eqnarray}%
A notable implication of Eq. $\left( \ref{33} \right)$ is:
\begin{equation}
\frac{M^{2}\nu _{\varphi }^{2}}{\Omega ^{3}}+\frac{\mathcal{D}}{\Omega }-2%
\sqrt{\mathcal{F}}-2=2n,  \label{39}
\end{equation}%
This implies that the energy eigenvalues adhere to a quantization condition. In contrast to Eqs. $\left( \ref{14} \right)$ and $\left( \ref{15} \right)$, we now have a fourth--order expression for the energy, as given by:
\begin{equation}
D_{4}\varepsilon ^{4}+D_{3}\varepsilon ^{3}+D_{2}\varepsilon
^{2}+D_{1}\varepsilon +D_{0}=0,  \label{40}
\end{equation}%
with%
\begin{eqnarray}
D_{4} &=&\frac{1}{\Omega ^{2}},  \label{41} \\
D_{3} &=&\frac{4M\omega a}{\Omega ^{2}\alpha },  \label{42} \\
D_{2} &=&\frac{2M^{2}\nu _{\varphi }^{2}}{\Omega ^{4}}-\frac{4\left(
n+1\right) }{\Omega }+\frac{2}{\Omega ^{2}}\left( L+2M^{2}\omega ^{2}\frac{%
a^{2}}{\alpha ^{2}}\right) ,  \label{43} \\
D_{1} &=&\left[ \frac{2M^{2}\nu _{\varphi }^{2}}{\Omega ^{4}}-\frac{4\left(
n+1\right) }{\Omega }+\frac{2L}{\Omega ^{2}}\right] 2M\omega \frac{a}{\alpha 
}-\frac{8al}{\hbar c\alpha ^{2}},  \label{44} \\
D_{0} &=&\frac{M^{2}\nu _{\varphi }^{2}}{\Omega ^{3}}\left[ \frac{M^{2}\nu
_{\varphi }^{2}}{\Omega ^{3}}-4\left( n+1\right) \right] +\left[ \frac{%
2M^{2}\nu _{\varphi }^{2}}{\Omega ^{3}}-4\left( n+1\right) +\frac{L}{\Omega }%
\right] \frac{L}{\Omega }  \notag \\
&&+4\left( n+1\right) ^{2}-\frac{4l^{2}}{\alpha ^{2}}-4\kappa _{\varphi
}^{2},  \label{45}
\end{eqnarray}%
with%
\begin{equation*}
L=2M\omega \frac{l}{\alpha }-M^{2}-2\kappa _{\varphi }\nu _{\varphi
}+k_{z}^{2}.
\end{equation*}%


\subsubsection{ Linear-like confinement $\left( \protect\kappa _{\protect%
\varphi }=0\right) $}

In this case, the Coulomb-type potential term is absent, and as a
consequence the scalar potential is reduced to the linear term in $\rho $.
Thus, the solutions are now given by%
\begin{equation}
H^{\left( 1\right) }\left( \rho \right) =c_{1}H_{b}\left( 2\sqrt{\mathcal{%
\Lambda }},\frac{2M\nu _{\varphi }}{\Omega ^{3/2}},\frac{M^{2}\nu _{\varphi
}^{2}}{\Omega ^{3}},\frac{\mathcal{\Delta }}{\Omega },0;\sqrt{\Omega }\rho
\right) ,
\end{equation}%
\begin{equation}
H^{\left( 2\right) }\left( z\right) =c_{2}\rho ^{-2\sqrt{\mathcal{\Lambda }}%
}H_{b}\left( -2\sqrt{\mathcal{\Lambda }},\frac{2M\nu _{\varphi }}{\Omega
^{3/2}},\frac{M^{2}\nu _{\varphi }^{2}}{\Omega ^{3}},\frac{\mathcal{\Delta }%
}{\Omega },0;\sqrt{\Omega }\rho \right) .
\end{equation}%
Again we discard the second solution because it diverges at $\rho =0$. The
condition to get polynomial solutions is now%
\begin{equation}
\frac{M^{2}\nu ^{2}}{\Omega ^{3}}+\frac{\mathcal{\Delta }}{\Omega }-2\sqrt{%
\mathcal{\Lambda }}-2=n.
\end{equation}%
As before, the above condition implies in the quantization of the energy
eigenvalues which is equivalent to Eq.$\left( \ref{40}\right) $, with the
coefficients given by $\left( \ref{41}\right) -\left( \ref{45}\right) $,
with $\kappa _{\varphi }=0$.


\section{Spinless particles}

We shall now investigate the relativistic Landau levels of a charged spinless particle in the spacetime of a magnetized rotating string, specifically one endowed with intrinsic magnetic flux $\Phi$, and devoid of any external electromagnetic field \cite{de2008berry,furtado2000harmonic}. The corresponding gauge coupling is derived by substituting $B\rightarrow B=\Phi /\alpha \pi \rho ^{2}$ into Eq. (\ref{4}). Consequently, in this scenario, the radial equation takes the form:
\begin{equation}
\rho ^{2}\frac{d^{2}R}{d\rho ^{2}}+\rho \frac{dR}{d\rho }+\left( \delta \rho
^{2}-\Sigma \right) R=0,  \label{48}
\end{equation}%
where $\delta $\ and $\Sigma $\  are %
\begin{eqnarray}
\Sigma &=&\left( \frac{l}{\alpha }+\frac{a}{\alpha }\varepsilon -\epsilon 
\frac{\Phi }{\alpha }\right) ^{2},  \label{49} \\
\delta &=&\varepsilon ^{2}-M^{2}-k_{z}^{2},  \label{50}
\end{eqnarray}%
where $\epsilon =e/2\pi \hbar c$.

The solutions to Eq. (\ref{48}) are expressed in terms of Bessel's functions of the first kind, denoted as $J_{\lambda }\left( z\right)$, and the second kind, represented by $Y_{\lambda }\left( z\right)$. Specifically:
\begin{equation}
R\left( \rho \right) =C_{1}J_{\sqrt{\Sigma }}\left( \sqrt{\delta }\rho
\right) +C_{2}Y_{\sqrt{\Sigma }}\left( \sqrt{\delta }\rho \right) ,
\end{equation}%
Here, $C_{1}$ and $C_{2}$ are constants. The function $J_{\lambda }\left(z\right)$ is non-zero at the origin when $\lambda =0$. On the other hand, $Y_{\sqrt{\Sigma }}$ is always divergent at the origin, so we discard it and consider $\lambda \neq 0$. It is noteworthy that when $\Phi =0$, we recover the wave function found in \cite{krori1994exact}. To determine the energy eigenvalues, we impose the so--called hard--wall condition. Under this boundary condition, the particle's wave function becomes zero at a specific radius $\rho = r_{w}$, which is arbitrarily chosen and situated far from the origin. Consequently, we can utilize the asymptotic expansion for large arguments of $J_{\lambda}\left(z\right)$, given by:
\begin{equation}
J_{\lambda }\left( z\right) \approx \sqrt{\frac{2}{\pi z}}\cos \left( z-%
\frac{\lambda \pi }{2}-\frac{\pi }{4}\right) ,
\end{equation}%
in a such way that%
\begin{equation}
\sqrt{\delta }r_{w}-\frac{\sqrt{\Sigma }\pi }{2}-\frac{\pi }{4}=\frac{\pi }{2%
}+n\pi ,  \label{53}
\end{equation}%
for $n\in 
\mathbb{Z}
$. After putting Eqs.$\left( \ref{49}\right) $ and $\left( \ref{50}\right) $
into $\left( \ref{53}\right) $, it reads%
\begin{equation}
r_{w}\sqrt{\varepsilon ^{2}-M^{2}-k_{z}^{2}}\mp \frac{\pi }{2}\left( \frac{l%
}{\alpha }+\frac{a}{\alpha }\varepsilon -\epsilon \frac{\Phi }{\alpha }%
\right) =\pi \left( \frac{3}{4}+n\right) ,  \label{54}
\end{equation}%
Here, the upper and lower signs are associated with $\left( \frac{l}{\alpha }+\frac{a}{\alpha }\varepsilon -\epsilon \frac{\Phi }{\alpha }\right) \leq 0$ or $\left( \frac{l}{\alpha }+\frac{a}{\alpha }\varepsilon -\epsilon \frac{\Phi }{\alpha }\right) >0$, respectively. Eq. (\ref{54}) can be reformulated as 
\begin{equation*}
A_{1}\varepsilon ^{2}+A_{2}\varepsilon +A_{3}=0, 
\end{equation*}%
being%
\begin{eqnarray}
A_{1} &=&r_{w}^{2}-\frac{a^{2}\pi ^{2}}{4\alpha ^{2}}, \\
A_{2} &=&-\frac{a\pi ^{2}}{2\alpha }\left[ \frac{l}{\alpha }-\epsilon \frac{%
\Phi }{\alpha }\pm \left( 2n+\frac{3}{2}\right) \right] , \\
A_{3} &=&-r_{w}^{2}\left( M^{2}+k_{z}^{2}\right) -\left( \frac{l}{\alpha }%
-\epsilon \frac{\Phi }{\alpha }\right) \frac{\pi ^{2}}{4}-\left( n+\frac{3}{4%
}\right) ^{2}\pi ^{2}  \notag \\
&&\mp \left( n+\frac{3}{4}\right) \left( \frac{l}{\alpha }-\epsilon \frac{%
\Phi }{\alpha }\right) \pi ^{2}.
\end{eqnarray}%
As $r_{w}$ becomes significantly large, $\varepsilon$ simplifies to:%
\begin{equation}
\varepsilon _{+}\approx \sqrt{M^{2}+k_{z}^{2}}+\frac{a\pi ^{2}}{4\alpha
r_{w}^{2}}\left[ \frac{l}{\alpha }-\epsilon \frac{\Phi }{\alpha }\pm \left(
2n+\frac{3}{2}\right) \right] ,
\end{equation}%
\begin{equation}
\varepsilon _{-}\approx -\sqrt{M^{2}+k_{z}^{2}}+\frac{a\pi ^{2}}{4\alpha
r_{w}^{2}}\left[ \frac{l}{\alpha }-\epsilon \frac{\Phi }{\alpha }\pm \left(
2n+\frac{3}{2}\right) \right] .
\end{equation}%

Now, let us examine $\varepsilon_{+}$ (where $\varepsilon_{+}=E_{+}/\hbar c$) and assume that $k_{z}\ll M$. Under this assumption, and given that $\frac{l}{\alpha} \geq \epsilon \frac{\Phi}{\alpha}$ (refer to Eq. (54)), we obtain:
\begin{equation}
E_{+}\approx mc^{2}+\frac{a\pi ^{2}\hbar c}{4\alpha r_{w}^{2}}\left[ \frac{l%
}{\alpha }-\epsilon \frac{\Phi }{\alpha }+2n+\frac{3}{2}\right].
\end{equation}%
This reveals that in the absence of rotation, the energy eigenvalues reduce to the rest energy of the particle, regardless of the parameter $\alpha$. In essence, the eigenenergies remain unchanged whether or not there is a (static) magnetized cosmic string in space. However, they undergo a splitting when the string is in rotation.


\section{Conclusions and remarks}

In this study, we generalized the spacetime induced by a rotating cosmic string, considering anisotropic effects due to the breaking of Lorentz violation. Specifically, we explored the energy levels of a massive spinless particle that was covariantly coupled to a uniform magnetic field aligned with the string. Following this, we introduced a scalar potential featuring both a Coulomb--type and a linear confining term, comprehensively solving the Klein–-Gordon equations for each configuration. Subsequently, by imposing rigid--wall boundary conditions, we determined the Landau levels when the linear defect itself possessed magnetization. Notably, our analysis unveiled the occurrence of Landau quantization even in the absence of gauge fields, provided the string possessed spin. Finally, the thermodynamic properties were computed in these scenarios.


\section*{Acknowledgments}
\hspace{0.5cm}

The authors also express their gratitude to FAPEMA, CNPq and CAPES (Brazilian research agencies) for invaluable financial support. In particular, L. Lisboa--Santos is supported by FAPEMA BPD-11962/22 and A. A. Araújo Filho is supported by Conselho Nacional de Desenvolvimento Cient\'{\i}fico e Tecnol\'{o}gico (CNPq) and Fundação de Apoio à Pesquisa do Estado da Paraíba (FAPESQ) -- [150891/2023-7].

\section{Data Availability Statement}

Data Availability Statement: No Data associated in the manuscript


\bibliographystyle{ieeetr}
\bibliography{main}

\begin{thebibliography}{10}

\bibitem{01}
A.~Vilenkin, ``Inflating horizons of particle astrophysics and cosmology,''
  2006.

\bibitem{02}
T.~W. Kibble, ``Cosmic strings reborn?,'' {\em arXiv preprint
  astro-ph/0410073}, 2004.

\bibitem{03}
E.~{\v{S}}im{\'a}nek, ``Gravitational field of a spinning sigma-model cosmic
  string,'' {\em Physical Review D}, vol.~78, no.~4, p.~045014, 2008.

\bibitem{04}
H.~Mota, E.~B. de~Mello, C.~Bessa, and V.~Bezerra, ``Light-cone fluctuations in
  the cosmic string spacetime,'' {\em Physical Review D}, vol.~94, no.~2,
  p.~024039, 2016.

\bibitem{05}
K.~Jusufi, ``Light deflection with torsion effects caused by a spinning cosmic
  string,'' {\em The European Physical Journal C}, vol.~76, no.~6, p.~332,
  2016.

\bibitem{06}
T.~Charnock, A.~Avgoustidis, E.~J. Copeland, and A.~Moss, ``Cmb constraints on
  cosmic strings and superstrings,'' {\em Physical Review D}, vol.~93, no.~12,
  p.~123503, 2016.

\bibitem{07}
M.~Salazar-Ram{\'\i}rez, D.~Ojeda-Guill{\'e}n, and R.~Mota, ``Algebraic
  approach and coherent states for a relativistic quantum particle in cosmic
  string spacetime,'' {\em Annals of Physics}, vol.~372, pp.~283--296, 2016.

\bibitem{08}
A.~Linde, ``Chaotic inflation in supergravity and cosmic string production,''
  {\em Physical Review D}, vol.~88, no.~12, p.~123503, 2013.

\bibitem{09}
E.~B. de~Mello, V.~B. Bezerra, and Y.~V. Grats, ``Self-forces in the spacetime
  of multiple cosmic strings,'' {\em Classical and Quantum Gravity}, vol.~15,
  no.~7, p.~1915, 1998.

\bibitem{10}
C.~Muniz and V.~Bezerra, ``Self-force on an electric dipole in the spacetime of
  a cosmic string,'' {\em Annals of Physics}, vol.~340, no.~1, pp.~87--93,
  2014.

\bibitem{11}
M.~Sazhin, O.~Khovanskaya, M.~Capaccioli, G.~Longo, M.~Paolillo, G.~Covone,
  N.~Grogin, and E.~Schreier, ``Gravitational lensing by cosmic strings: what
  we learn from the csl-1 case,'' {\em Monthly Notices of the Royal
  Astronomical Society}, vol.~376, no.~4, pp.~1731--1739, 2007.

\bibitem{12}
V.~Bezerra, V.~Mostepanenko, and R.~T. Filho, ``Particle creation in the chiral
  cosmic string spacetime,'' {\em International Journal of Modern Physics D},
  vol.~11, no.~03, pp.~437--445, 2002.

\bibitem{13}
V.~De~Lorenci, R.~De~Paola, and N.~Svaiter, ``From spinning to non-spinning
  cosmic string spacetime,'' {\em Classical and Quantum Gravity}, vol.~16,
  no.~10, p.~3047, 1999.

\bibitem{14}
V.~De~Lorenci, R.~De~Paola, and N.~Svaiter, ``From spinning to non-spinning
  cosmic string spacetime,'' {\em Classical and Quantum Gravity}, vol.~16,
  no.~10, p.~3047, 1999.

\bibitem{15}
S.~Deser, R.~Jackiw, and G.~Hooft, ``Three-dimensional einstein gravity:
  dynamics of flat space,'' {\em Annals of Physics}, vol.~152, no.~1,
  pp.~220--235, 1984.

\bibitem{16}
J.~R. Gott and M.~Alpert, ``General relativity in a (2+ 1)-dimensional
  space-time,'' {\em General Relativity and Gravitation}, vol.~16,
  pp.~243--247, 1984.

\bibitem{17}
A.~Barros, V.~Bezerra, and C.~Romero, ``Global aspects of gravitomagnetism,''
  {\em Modern Physics Letters A}, vol.~18, no.~37, pp.~2673--2679, 2003.

\bibitem{18}
V.~Bezerra, ``Some remarks on loop variables, holonomy transformation, and
  gravitational aharonov-bohm effect,'' {\em Annals of Physics}, vol.~203,
  no.~2, pp.~392--409, 1990.

\bibitem{20}
B.~Jensen and H.~H. Soleng, ``General-relativistic model of a spinning cosmic
  string,'' {\em Physical Review D}, vol.~45, no.~10, p.~3528, 1992.

\bibitem{21}
H.~H. Soleng, ``Negative energy densities in extended sources generating closed
  timelike curves in general relativity with and without torsion,'' {\em
  Physical Review D}, vol.~49, no.~2, p.~1124, 1994.

\bibitem{22}
N.~{\"O}zdemir, ``Spinning cosmic strings: A general class of solutions,'' {\em
  International Journal of Modern Physics A}, vol.~20, no.~13, pp.~2821--2832,
  2005.

\bibitem{23}
L.~de~Andrade, ``Cosmic strings and closed time-like curves in teleparallel
  gravity,'' {\em arXiv preprint gr-qc/0102094}, 2001.

\bibitem{24}
R.~J. Slagter, ``Time evolution and matching conditions of spinning gauge
  strings,'' {\em Physical Review D}, vol.~54, no.~8, p.~4873, 1996.

\bibitem{25}
A.~Mostafazadeh, ``Relativistic adiabatic approximation and geometric phase,''
  {\em Journal of Physics A: Mathematical and General}, vol.~31, no.~38,
  p.~7829, 1998.

\bibitem{26}
M.~Cunha, C.~Muniz, H.~Christiansen, and V.~Bezerra, ``Relativistic landau
  levels in the rotating cosmic string spacetime,'' {\em The European Physical
  Journal C}, vol.~76, pp.~1--7, 2016.

\bibitem{27}
G.~de~A~Marques, C.~Furtado, V.~B. Bezerra, and F.~Moraes, ``Landau levels in
  the presence of topological defects,'' {\em Journal of Physics A:
  Mathematical and General}, vol.~34, no.~30, p.~5945, 2001.

\bibitem{28}
K.~Bakke, L.~Ribeiro, C.~Furtado, and J.~Nascimento, ``Landau quantization for
  a neutral particle in the presence of topological defects,'' {\em Physical
  Review D}, vol.~79, no.~2, p.~024008, 2009.

\bibitem{29}
K.~Bakke, ``Noninertial effects on a dirac neutral particle inducing an
  analogue of the landau quantization in the cosmic string spacetime,'' {\em
  Brazilian Journal of Physics}, vol.~42, pp.~437--444, 2012.

\bibitem{30}
E.~Figueiredo~Medeiros and E.~Bezerra~de Mello, ``Relativistic quantum dynamics
  of a charged particle in cosmic string spacetime in the presence of magnetic
  field and scalar potential,'' {\em The European Physical Journal C}, vol.~72,
  no.~6, p.~2051, 2012.

\bibitem{31}
M.~Bueno, C.~Furtado, and A.~de~M.~Carvalho, ``Landau levels in graphene layers
  with topological defects,'' {\em The European Physical Journal B}, vol.~85,
  pp.~1--5, 2012.

\bibitem{kostelecky2011data}
V.~A. Kosteleck{\`y} and N.~Russell, ``Data tables for lorentz and c p t
  violation,'' {\em Reviews of Modern Physics}, vol.~83, no.~1, p.~11, 2011.

\bibitem{colladay2004statistical}
D.~Colladay and P.~McDonald, ``Statistical mechanics and lorentz violation,''
  {\em Physical Review D}, vol.~70, no.~12, p.~125007, 2004.

\bibitem{casana2008lorentz}
R.~Casana, M.~M. Ferreira~Jr, and J.~S. Rodrigues, ``Lorentz-violating
  contributions of the carroll-field-jackiw model to the cmb anisotropy,'' {\em
  Physical Review D}, vol.~78, no.~12, p.~125013, 2008.

\bibitem{casana2009finite}
R.~Casana, M.~M. Ferreira~Jr, J.~S. Rodrigues, and M.~R. Silva, ``Finite
  temperature behavior of the c p t-even and parity-even electrodynamics of the
  standard model extension,'' {\em Physical Review D}, vol.~80, no.~8,
  p.~085026, 2009.

\bibitem{gomes2010free}
M.~Gomes, T.~Mariz, J.~Nascimento, A.~Petrov, A.~Santos, and A.~da~Silva,
  ``Free energy of lorentz-violating qed at high temperature,'' {\em Physical
  Review D}, vol.~81, no.~4, p.~045013, 2010.

\bibitem{aa2020lorentz}
A.~A. Ara{\'u}jo~Filho, ``Lorentz-violating scenarios in a thermal reservoir,''
  {\em The European Physical Journal Plus}, vol.~136, no.~4, pp.~1--14, 2021.

\bibitem{araujo2022particles}
A.~A. Ara{\'u}jo~Filho, ``Particles in loop quantum gravity formalism: a
  thermodynamical description,'' {\em Annalen der Physik}, vol.~534, no.~12,
  p.~2200383, 2022.

\bibitem{araujo2021bouncing}
A.~A. Ara{\'u}jo~Filho and A.~Y. Petrov, ``Bouncing universe in a heat bath,''
  {\em International Journal of Modern Physics A}, vol.~36, no.~34n35,
  p.~2150242, 2021.

\bibitem{maluf2020thermodynamic}
A.~A. Ara{\'u}jo~Filho and R.~Maluf, ``Thermodynamic properties in
  higher-derivative electrodynamics,'' {\em Brazilian Journal of Physics},
  vol.~51, no.~3, pp.~820--830, 2021.

\bibitem{reis2020thermal}
A.~A. Ara{\'u}jo~Filho and J.~Reis, ``Thermal aspects of interacting quantum
  gases in lorentz-violating scenarios,'' {\em The European Physical Journal
  Plus}, vol.~136, no.~3, pp.~1--30, 2021.

\bibitem{das2009relativistic}
S.~Das, S.~Ghosh, and D.~Roychowdhury, ``Relativistic thermodynamics with an
  invariant energy scale,'' {\em Physical Review D}, vol.~80, no.~12,
  p.~125036, 2009.

\bibitem{araujo2021higher}
A.~A. Ara{\'u}jo~Filho and A.~Y. Petrov, ``Higher-derivative lorentz-breaking
  dispersion relations: a thermal description,'' {\em The European Physical
  Journal C}, vol.~81, no.~9, p.~843, 2021.

\bibitem{araujo2022thermal}
A.~A. Ara{\'u}jo~Filho, {\em Thermal aspects of field theories}.
\newblock Amazon. com, 2022.

\bibitem{araujo2023thermodynamics}
A.~A. Ara{\'u}jo~Filho, H.~Hassanabadi, J.~Reis, and L.~Lisboa-Santos,
  ``Thermodynamics of a quantum ring modified by lorentz violation,'' {\em
  Physica Scripta}, vol.~98, no.~6, p.~065943, 2023.

\bibitem{mazur1986spinning}
P.~O. Mazur, ``Spinning cosmic strings and quantization of energy,'' {\em
  Physical review letters}, vol.~57, no.~8, p.~929, 1986.

\bibitem{cunha2016relativistic}
M.~Cunha, C.~Muniz, H.~Christiansen, and V.~Bezerra, ``Relativistic landau
  levels in the rotating cosmic string spacetime,'' {\em The European Physical
  Journal C}, vol.~76, pp.~1--7, 2016.

\bibitem{figueiredo2012relativistic}
E.~Figueiredo~Medeiros and E.~Bezerra~de Mello, ``Relativistic quantum dynamics
  of a charged particle in cosmic string spacetime in the presence of magnetic
  field and scalar potential,'' {\em The European Physical Journal C}, vol.~72,
  no.~6, p.~2051, 2012.

\bibitem{de2001landau}
G.~de~A~Marques, C.~Furtado, V.~B. Bezerra, and F.~Moraes, ``Landau levels in
  the presence of topological defects,'' {\em Journal of Physics A:
  Mathematical and General}, vol.~34, no.~30, p.~5945, 2001.

\bibitem{aa}
P.~Sedaghatnia, H.~Hassanabadi, J.~Porf{\'\i}rio, W.~Chung, {\em et~al.},
  ``Thermodynamical properties of a deformed schwarzschild black hole via dunkl
  generalization,'' {\em arXiv preprint arXiv:2302.11460}, 2023.

\bibitem{bb}
N.~Heidari, H.~Hassanabadi, A.~A. Ara{\'u}jo~Filho, J, S.~Zare, and
  P.~Porf{\'\i}rio, ``Gravitational signatures of a non--commutative stable
  black hole,'' {\em Physics of the Dark Universe}, p.~101382, 2023.

\bibitem{cc}
J.~Furtado, H.~Hassanabadi, J.~Reis, {\em et~al.}, ``Thermal analysis of
  photon-like particles in rainbow gravity,'' {\em arXiv preprint
  arXiv:2305.08587}, 2023.

\bibitem{dd}
A.~A. Ara{\'u}jo~Filho, J.~Furtado, J.~Reis, and J.~Silva, ``Thermodynamical
  properties of an ideal gas in a traversable wormhole,'' {\em Classical and
  Quantum Gravity}, vol.~40, no.~24, p.~245001, 2023.

\bibitem{vieira2015quantum}
H.~S. Vieira and V.~B. Bezerra, ``Quantum newtonian cosmology and the
  biconfluent heun functions,'' {\em Journal of Mathematical Physics}, vol.~56,
  no.~9, 2015.

\bibitem{de2008berry}
J.~de~S.~Carvalho, E.~Passos, C.~Furtado, and F.~Moraes, ``Berry’s phase for
  a spin 1/2 particle in the presence of topological defects,'' {\em The
  European Physical Journal C}, vol.~57, pp.~817--822, 2008.

\bibitem{furtado2000harmonic}
C.~Furtado and F.~Moraes, ``Harmonic oscillator interacting with conical
  singularities,'' {\em Journal of Physics A: Mathematical and General},
  vol.~33, no.~31, p.~5513, 2000.

\bibitem{krori1994exact}
K.~Krori, P.~Borgohain, and D.~Das, ``Exact scalar and spinor solutions in the
  field of a stationary cosmic string,'' {\em Journal of Mathematical Physics},
  vol.~35, no.~2, pp.~1032--1036, 1994.

\end{thebibliography}

\end{document}